# Dynamic Separation of Chaotic Signals in the Presence of Noise

Yuri V. Andreyev [a], Alexander S. Dmitriev [a] and Elena V. Efremova [b]

[a] *Institute of Radioengineering and Electronics, Russian Academy of Sciences, Mokhovaya st. 11/7, GSP-9, Moscow, K-9, 101999 Russia*

[b] *Moscow Institute for Physics and Technology, Institutskii st. 9, Dolgoprudnyi, Moscow region, 141700 Russia*

**Abstract**

The problem of separation of an observed sum of chaotic signals into the individual components in the presence of noise on the path to the observer is considered. A noise threshold is found above which high-quality separation is impossible. Below the threshold, each signal is recovered with any prescribed accuracy. This effect is shown to be associated with the information content of the chaotic signals and a theoretical estimate is given for the threshold.

Keywords: dynamic chaos, signal separation, effect of noise on chaotic systems, chaotic signal processing, information content of chaotic signal, denoising

PACS numbers:   05.45.Gg,   05.45.Vx,   05.45.Xt,   07.50.Qx

# I. Introduction

Chaotic oscillators exist in many physical, biological, electronic, mechanical, and other systems. Oscillations produced by such sources are often analyzed with the use of observations of a single component of the process. Besides, the observed signal can represent not the pure component of the process, but its certain transformation, or it can be corrupted by uncontrollable distortions. Typical problems that an observer confronts by analyzing chaotic sources are cleaning chaotic signals off noise [1–6], reconstructing the chaotic attractor [7–9], and estimating its correlation dimension [10–12].

The situation becomes more complicated when there are two or more chaotic sources and the observer receives a certain combination of their signals. The simplest case is the sum of the chaotic signals. In order to analyze each source, it is necessary to separate the signals from the observed sum into the individual components. Can this problem be solved in the case of a single observable?

As was shown in [13], this was possible, in principle, in the case that the observer knew the equations describing the dynamic systems. The authors used the principle of chaotic synchronization to demonstrate this possibility. Unfortunately, the method using chaotic synchronization was very sensitive to external noise.

In this paper we use a different approach based on iteration of chaotic systems in reverse time. We discuss this approach on example of two chaotic sources represented by two maps of logistic parabola with different parameters. On this simple example we show that chaotic signals can be separated not only in the absence but also in the presence of noise. As is found, there is a certain threshold of the noise value above which high-grade separation becomes impossible. Below the threshold, each signal can be recovered with any prescribed (high) accuracy. We discuss this effect and determine that it is coupled not with the concrete method of separation, but is explained by basic reasons associated with the information content of the chaotic signals and with impossibility for an observer to receive this information content without serious distortions in the presence of noise stronger than a certain threshold.

We also give a theoretical estimate for the threshold noise value and compare the threshold obtained numerically with the theoretical estimate. The difference between the two threshold values is a measure of the concrete algorithm efficiency. The less the difference, the more efficient is the algorithm. Then we introduce a "mutibranch" algorithm for chaotic signal separation and demonstrate that it can provide separation efficiency close to the theoretical estimate.

Finally, we discuss to what extent the obtained results can be generalized to other chaotic sources.

## 2. Separation Method

The problem we consider here is as follows. Let there be 2 chaotic sources producing chaotic signals $x_j(k)$, $j = 1, 2$; and $k$ be discrete time. On the path (a channel) to an observer, the signals $x_j(k)$ are summed. In general, the sum signal is also contaminated by an additive noise $\eta(k)$ (Fig. 1). The observer has to separate the individual signals from the sum.

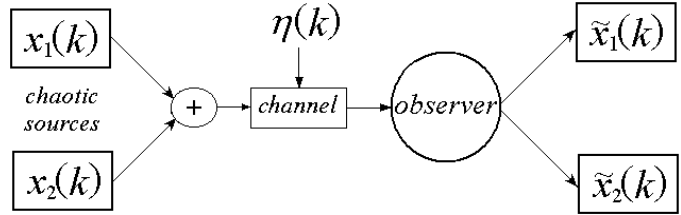

Fig. 1. Separation of chaotic signals.

Here we consider the problem of chaotic signal separation on example of a single channel connecting two chaotic sources with the observer. Let the sources be chaotic oscillators represented by similar maps $x(k+1) = f(x(k), \mu)$ with different parameters $\mu$, and the map functions be denoted by $f_1(x) = f(x, \mu_1)$ and $f_2(x) = f(x, \mu_2)$. The dynamics equations are

$$x_1(k+1) = f_1(x_1(k)),$$
$$x_2(k+1) = f_2(x_2(k)), \qquad (1)$$

where $k$ is discrete time.

The signal in the channel is

$$u(k) = x_1(k) + x_2(k) + \eta(k). \qquad (2)$$

So, the problem can be rigorously defined as follows. Given a sequence of samples of a sum signal $\{u(k)\}$, $k = 1, 2, …, N$; knowing the dynamics of the systems generating the chaotic signals (here, the functions $f_1$ and $f_2$), and given (good) estimates $\tilde{x}_1(N)$ and $\tilde{x}_2(N)$ at $N$th time moment; to obtain estimates $\tilde{x}_1(k)$ and $\tilde{x}_2(k)$, $k = 1, 2, …, N$, of the oscillator signals $x_1(k)$ and $x_2(k)$ on the entire time interval, satisfying the dynamics of sources (1) and the most close to $x_1(k)$ and $x_2(k)$, respectively.

Let for certainty the chaotic sources be described by the maps of logistic parabola $f(x) = \mu x(1-x)$

$$x_1(k+1) = \mu_1 x_1(k)(1 - x_1(k)),$$
$$x_2(k+1) = \mu_2 x_2(k)(1 - x_2(k)). \qquad (3)$$

The idea of the proposed separation method is as follows. The observer has the maps $f^{-1}$ inverse to those that generate the chaotic signals (Fig. 1):

$$x_1(k-1) = f_1^{-1}(x_1(k)),$$
$$x_2(k-1) = f_2^{-1}(x_2(k)). \qquad (4)$$

Iteration of maps (4) is equivalent to backward iteration of equations (1). The sum signal $u(k)$ (2) comes from the channel. Let the observer at $k$th moment have a sample of the sum of chaotic signals $u(k)$ and separate estimates of the values of the chaotic signals of both oscillators, i.e., an estimate $\tilde{x}_1(k)$ for $x_1(k)$ and $\tilde{x}_2(k)$ for $x_2(k)$. Iteration of the maps of inverse systems (4) with initial conditions $\tilde{x}_1(k)$ and $\tilde{x}_2(k)$ gives estimates of the signals at $(k-1)$th moment (Fig. 2)

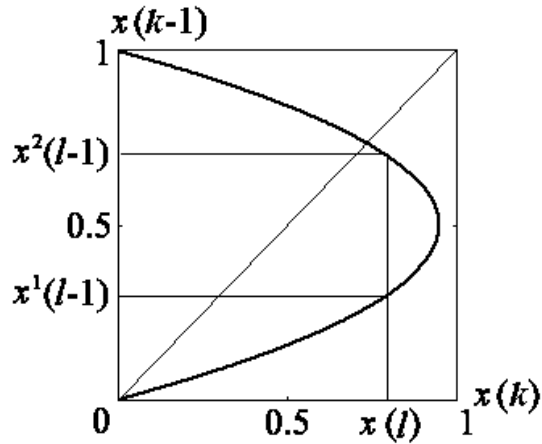

Fig. 2. Two-valued function $f^{-1}(.)$ of the map inverse to logistic map.

Since maps (3) are stretching on the average (over the attractors), inverse maps (4) are contracting (on the average). Hence, the estimates for signals $x_1$ and $x_2$ at $(k-1)$th moment, obtained from the estimates $\tilde{x}_1(k)$ and $\tilde{x}_2(k)$, will on the average be more accurate that the initial estimates $\tilde{x}_1(k)$ and $\tilde{x}_2(k)$. However, maps (4) are two-valued and each iteration of (4) gives two values for a single argument: two potential estimates $\tilde{x}_1^1(k-1)$ and $\tilde{x}_1^2(k-1)$ for the first source, and two estimates $\tilde{x}_2^1(k-1)$ and $\tilde{x}_2^2(k-1)$ for the second source. So, we have to choose the "proper" branch of each map function by the iteration. This can be organized as follows. These two pairs of two estimates give us four possible combinations for the sum signal estimates at $(k-1)$th moment:

$$u_{ij}(k-1) = \tilde{x}_1^i(k-1) + \tilde{x}_2^j(k-1), \quad i,j = 1, 2. \qquad (5)$$

At the same time, we know that the signal that came to the observer from the channel at $(k-1)$th moment was $u(k-1)$. We can make the proper choice if we compare the value of $u(k-1)$ with those of $u_{ij}(k-1)$. Indeed, the best choice $(i,j)$ is the combination of the branches that minimizes deviation of the estimates of the sum of two chaotic signals from the real sum signal that came from the channel at that moment:

$$(i, j): |u(k-1) - u_{ij}(k-1)| = \min_{p,q} |u(k-1) - u_{pq}(k-1)|, \quad p, q = 1, 2. \qquad (6)$$

The scheme for the choice of proper branch combinations at $(k-1)$th and other moments is illustrated in Fig. 3. The values of the channel signal $u(l)$, $l < k$, are denoted by asterisks. At $l$th moment from four possible values of $u_{ij}(l)$ we take the one most close to $u(l)$, the same is done at $(l-1)$th moment, and so on. Thus, successive application of the discussed procedure allows us to separate the signals on the entire time interval $(1, N)$.

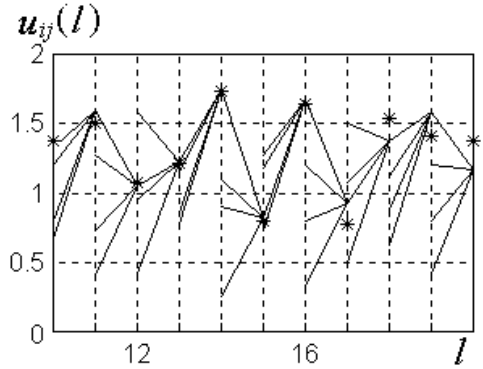

Fig. 3. Choice between the sum signal branches, produced by iteration of inverse maps.
$l$ is discrete time, $u_{ij}(l) = \tilde{x}_1^i(l) + \tilde{x}_2^j(l)$, $i, j = 1, 2$.
The values of the sum signal $u(l)$ are denoted by asterisks.

If $\lambda$ is Lyapunov exponent of a map (averaged over the map attractor), then the average stretching factor of the map is $e^{\lambda}$, and the inverse map contraction factor is $e^{-\lambda}$. So, the estimate errors $\delta_1(l)$ and $\delta_2(l)$ of the separated signals $\tilde{x}_1(l)$ and $\tilde{x}_2(l)$

$$\delta_1(l) = |\tilde{x}_1(l) - x_1(l)|,$$
$$\delta_2(l) = |\tilde{x}_2(l) - x_2(l)| \qquad (7)$$

decrease exponentially (on the average)

$$\delta_1(l) = \delta_1(N) \cdot \exp(-\lambda_1(N - l)),$$
$$\delta_2(l) = \delta_2(N) \cdot \exp(-\lambda_2(N - l)), \qquad (8)$$

where $\delta_1(N) = |\tilde{x}_1(N) - x_1(N)|$ and $\delta_2(N) = |\tilde{x}_2(N) - x_2(N)|$ are initial estimate errors, and $\lambda_1$ and $\lambda_2$ are Lyapunov exponents of the trajectories of the first and the second systems, respectively.

In agreement with expression (8), the closeness of the signals $\tilde{x}_1$ and $\tilde{x}_2$ recovered by the observer to the signals $x_1$ and $x_2$ of the sources improves exponentially with each step of inverse function it-eration and eventually achieves the limit of calculation accuracy. In our numerical experiments with accuracy $\varepsilon$ (in the case of double-precision arithmetic $\varepsilon \sim 10^{-16}$ (–320 dB), or 16 significant digits), the limit attainable closeness, i.e., separation accuracy, is achieved after $p = -\log(\varepsilon/\delta)/\lambda$ steps at most, where $\delta$ is the initial estimate error. Thus, the described procedure allows one to separate the signals, given on interval $(1, N)$, nearly on the entire interval $(1, N–p)$ as accurately as the machine arithmetic, with a few less accurate samples at the end of the separated signal sequences. This less accuracy of the ending $p$ samples can be explained by the lack of information necessary to separate the signals on the interval $(N–p, N)$.

Above, we discussed the method for signal separation under condition that estimates $\tilde{x}_1(N)$ and $\tilde{x}_2(N)$ of the chaotic sources' signals are known at $N$th moment. If the estimates are good, i.e., the initial error is of the order of the machine accuracy $\delta_{1,2}(N) \approx \varepsilon$, then $p = 0$, and the signals are separated as accurately as possible on the entire interval $(1, N)$. However, in general, we can take any pair of points $\tilde{x}_1(N)$ and $\tilde{x}_2(N)$ on the attractors of maps (3) as the initial estimates. Having started from these initial conditions the calculated trajectories of systems (4) converge with time to the trajectories of the chaotic oscillators. It is the time of convergence $p$ that depends on the particular choice of the initial points.

## 3. Verification of the Method and Efficiency Measures

The efficiency of the proposed method for chaotic signal separation was investigated on example of the sources described by logistic maps (3) with the parameters set at $\mu_1 = 3.7$ and $\mu_2 = 3.8$ (Lyapunov exponents $\lambda_1 = +0.355$ and $\lambda_2 = +0.432$).

In the absence of noise, the signals of the two logistic maps are efficiently separated. In computer simulation with double precision arithmetic the limit attainable accuracy of $\varepsilon \approx 10^{-16}$ in the case of the worst initial estimates is achieved at $\lambda = 0.355$ after $p = -\log(10^{-16})/\lambda \approx 100$ steps, while rather good practical accuracy of $10^{-3}$ is achieved already after 20 steps.

Since the method for chaotic signal separation proved to be applicable in the absence of noise, we concentrated our further investigation on the method resistance with respect to the channel noise. Gaussian, normally distributed noise $\eta(k)$ with variance $\sigma$ was added in the channel to the sources' signals $u(k) = x_1(k) + x_2(k) + \eta(k)$, $k = 1, ..., N$. The calculated signal recovery errors $x_i(k) – \tilde{x}_i(k)$, $i = 1, 2$, was treated as a residual noise in the separated signals. The noise levels (or inversely, sig-nal-to-noise ratio, $\text{SNR}_S$) for the separated signals were calculated as functions of the signal-to-noise ratio of the channel signal $u(k)$,

$$\text{SNR}_C = \langle(x_1 + x_2)^2\rangle/\sigma^2, \qquad (9)$$

or

$$\text{SNR}_{C,\,dB} = 10 \log[\langle(x_1 + x_2)^2\rangle/\sigma^2], \qquad (10)$$

(all signals were normalized to have zero mean values).

In order to quantitatively estimate the performance of the separation method, we need a measure of the separation efficiency. That is, in what case the signals can be considered separated, provided that the separated signals do not coincide with those of the sources? We considered several efficiency measures. The first is the difference $K$, dB, between the signal-to-noise ratios of the separated signals and of the channel signal $u(k)$

$$K = \text{SNR}_{S,\,dB} - \text{SNR}_{C,\,dB}. \qquad (11)$$

Separation is considered effective when the noise in the separated signals is lower than the channel noise, i.e., $K > 0$, and ineffective if the residual noise in the output signal is much higher than that in the channel ($K \ll 0$). We consider the threshold value of the channel $\text{SNR}_{C,\,dB}$ that gives $K = 0$ as the separation boundary.

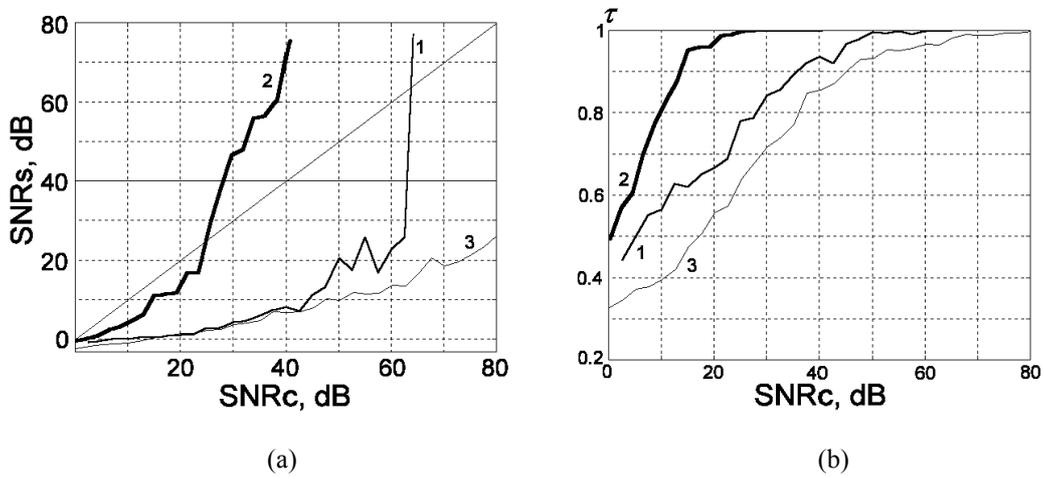

Fig. 4. (a) Signal-to-noise ratio of the separated signals, $\text{SNR}_S$, and (b) relative separation time $\tau$ as functions of noise in the channel, $\text{SNR}_C$. Curves are shown for (1) single-branch algorithm, (2) algorithm with 16 branches, and (3) method of [13]. The results were obtained with 10,000-sample chaotic sequences.

Calculation results are presented in Fig. 4. Our results are represented by curve 1, which shows that the region of effective separation extends to $\text{SNR}_C$ of ~ 65 dB. Given for comparison is curve 3, corresponding to results for the method of chaotic synchronization [13]. As can be seen in Fig. 4a, the noise in the signals separated by the method of chaotic synchronization is always higher than the channel noise and criterion (11) $K > 0$ is never fulfilled. Note that to the right of $SNR_C \approx 65$ dB, the noise in the separated signals rapidly decreases according to relations (8), and its value is eventually determined by only the number of backward iteration steps and by the machine calculation accuracy. This means that the signals are not only separated but also cleaned off noise.

In the region of $SNR_C < 65$ dB in the process of separation, sporadic separation error bursts can occur (Fig. 5). Even a single and very short burst can seriously spoil the $SNR_S$ value. However, if the channel noise is relatively small, the bursts occur seldom and most of the time the chaotic signals are well separated. Therefore, we also used another efficiency measure, the relative time $\tau$ of effective signal separation, which is defined as a fraction of the total time within which the local signal estimation error $\delta_{1,2} = |x_{1,2}(k) - \tilde{x}_{1,2}(k)|$ is less than $\delta_{1,2} < 0.01$, or $-40$ dB (Fig. 4b). Evidently, separation is effective if $\tau$ is close to one. As can be seen in Figs. 4a and 4b, both measures give close values of the separation boundaries.

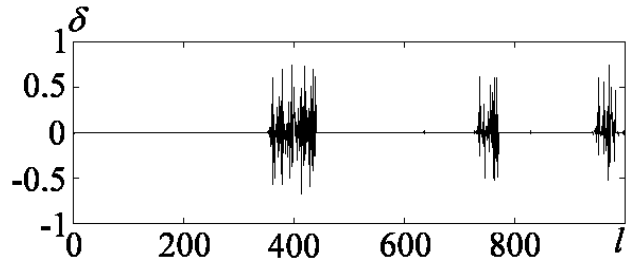

Fig. 5. Bursts of the signal separation error $\delta = x_1 - \tilde{x}_1$ for the single-branch algorithm; channel $SNR_C \approx 40$ dB; $l$ is discrete time.

Another efficiency measure may be the value of noise added to the channel signal (or inversely, $SNR_C$) at which the method provides a certain prescribed value of the separated signal $SNR_S$. We used the value of 40 dB SNR of the output recovered signals, since it is quite a good accuracy of the signal recovery (rms signal error of the order of 0.01). The above method gives 40 dB $SNR_S$ at $SNR_C = 60$ dB.

Simulation of the separation procedure shows that with increasing external noise the rate of the error bursts also increases, which gradually ruins the method efficiency. Analysis of the recovered signal waveforms shows that strong channel noise $\eta(k)$ at a particular moment can considerably shift the actual sum of the sources' signals $u(k) = x_1(k) + x_2(k) + \eta(k)$, and result in error bursts due to a wrong choice among the inverse map branches at that moment (see relation (6)). This wrong choice is exhibited on the next step of (4) map iteration as a sharp burst of separation error. Then the separated trajectories again begin to converge to the sources' signals, which can take a number of steps. These irregular "error" bursts are the reason for the residual noise $SNR_S$.

# 4. Threshold Effect and its Nature

As can be seen in Fig. 4, at the threshold value of $SNR_C \approx 63$ dB the signal-to-noise ratio of the separated signals $SNR_S$ jumps up by more than 40 dB. Is the presence of such a threshold a property of the discussed separation method, or a common feature of chaotic signal separation? And a more general question: are there principle limitations on the separation of chaotic signals and what are the reasons for these limitations?

To answer to these questions, let us consider the information properties of chaotic signals. A one-dimensional chaotic map generates Kolmogorov entropy with a mean rate equal to Lyapunov exponent $\lambda$ [8]. In the discussed case, Kolmogorov entropy is equivalent to information $I$, which is used, however, to be expressed in bits per iteration. Thus, chaotic 1-D map generates

$$I = \lambda/\ln 2 \qquad (12)$$

information bits per iteration. For example, logistic map with parameters $\mu_1 = 3.7$ and $\mu_2 = 3.8$ generates $I_1 = 0.51$ and $I_2 = 0.62$ bits per iteration on the average. These are mean values, however, the amount of generated information differs from iteration to iteration (Fig. 6a, b).

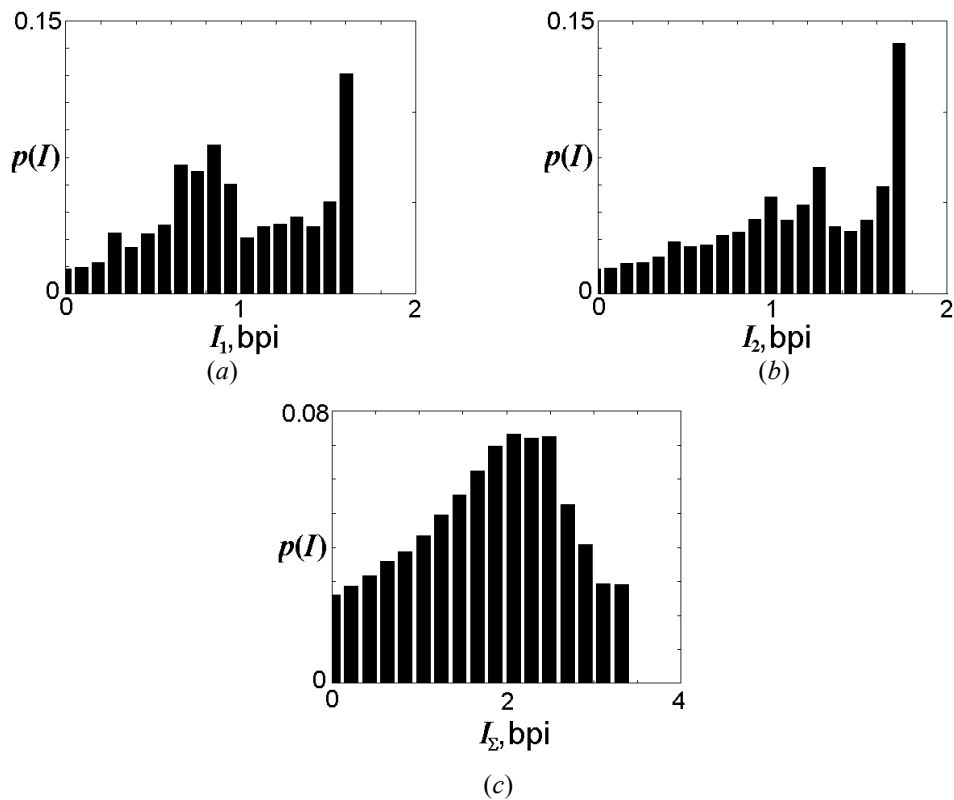

Fig. 6. Per-iteration distribution density of information produced by logistic map with the parameter set at (a) $\mu_1 = 3.7$; at (b) $\mu_2 = 3.8$; and (c) of the sum of the two signals.

According to expression (2) the observer receives the sum of chaotic signals $x_1$ и $x_2$ distorted by noise $\eta$. On each iteration step the sum of signals contains certain amount of information whose distribution density is presented in Fig. 6c. In order to separate the signals $x_1$ and $x_2$, it is necessary that the information is not lost due to contamination of the signal sum by the noise $\eta$. Note that one can treat expression (2) as a model of a "communication channel" with Gaussian noise through which a signal $x(k) = x_1(k) + x_2(k)$ is transmitted. According to Shannon theorem [14], the information-carrying capacity of the channel per iteration is equal to

$$\tilde{N} = \frac{1}{2}\log_2\left(1 + \frac{\langle x^2(k)\rangle}{\langle \eta^2(k)\rangle}\right) = \frac{1}{2}\log_2(1 + \text{SNR}_C). \qquad (13)$$

Maximum amount of information going through this channel is determined by the right boundary of the distribution density $I_{max}$ in Fig. 6c. This gives a necessary condition for the signal separation

$$C > I_{max}, \qquad (14)$$

consequently,

$$\text{SNR}_C > 2^{2I_{max}} - 1 \qquad (15)$$

or

$$\text{SNR}_{C,dB} = 10\ \lg(2^{2I_{max}} - 1). \qquad (16)$$

In the discussed case, $I_{max} \approx 3.4$ bits per iteration, hence

$$\text{SNR}_{C,dB} > 20\ \text{dB}. \qquad (17)$$

Comparison of the obtained estimate (17) with curve 1 in Fig. 4 indicates that the difference between the theoretical value of $\text{SNR}_{C,dB}$ and its value obtained with the above separation method is greater than 40 dB.

## 5. Multi-Branch Method

In the algorithm that was used above, the decision on which branch to take was made locally, in one point of time domain, and the preceding and the following histories were not taken into account. So, we developed and investigated a modified method whose efficiency is improved due to the use of nonlocal information at each iteration. We backtrack several branches simultaneously besides the one, optimal in the sense of condition (6), and choose among them by means of minimizing the deviation signal averaged over a certain time interval.

To do this, we build a tree of possible trajectories of $\tilde{x}_1$ and $\tilde{x}_2$ on the given interval $(l, k)$ and take the pair that minimizes the functional $G$ on this interval $(l, k)$

$$G = \sum_{i=l}^{k} \left(u(i) - (\tilde{x}_1(i) + \tilde{x}_2(i))\right)^2 . \qquad (18)$$

Due to evident inevitable restrictions on computational capabilities, memory resources, etc., we restrict the number of the backtracked branches, say to $M$, "best" in a certain sense, by means of discarding the least probable ones. Besides, specific dynamics of chaotic systems is to be taken into account: since the backtracked branches tend to converge due to relation (8), from time to time we remove the "stuck" ones in order to keep branches different. When the entire interval $(1, N)$ is processed, the separated signals $\tilde{x}_1$ and $\tilde{x}_2$ are obtained with the condition of the minimum $G$.

The results of separating chaotic signals with this algorithm are presented in Figs. 4a and 4b (curve 2). Sixteen branches were tracked back ($M = 16$). The results indicate that with this algorithm the chaotic signals are separated at the channel noise of about 25–30 dB, which is 35–40 dB better than with the algorithm with single branch and much closer to the theoretical separation limit.

## 6. Conclusions

In this paper we considered the problem of separation of an observed sum of chaotic signals into the individual components. It was shown that the problem could be solved not only in the absence but also in the presence of noise on the path to the observer. A certain noise threshold was found above which high-grade separation became impossible. Below the threshold, each signal was recovered with a prescribed accuracy. We proved that this effect was associated with the information content of the chaotic signals and gave a theoretical estimate for the threshold which was the limit for separation of chaotic signals in the presence of noise.

Two different implementations of the method for separation were proposed and verified on example of a model of two chaotic sources represented by two logistic maps with different parameters. A simple single-branch algorithm allowed us to separate chaotic signals at the channel $SNR_C > 60–65$ dB, and an advanced algorithm using several backtracked branches gave efficient separation of the signals at $SNR_C$ as low as 25–30 dB, while theoretical estimate for the separation threshold was about 20 dB.

Though we discussed here the problem of separation of chaotic signals in the case of one-dimensional systems, the above estimation of the separation limits is valid for a broader class of chaotic sources. The separation methods can be directly generalized to the case of $m > 2$ chaotic sources (represented by 1-D systems), and also to the sources described by multidimensional hyperbolic maps that have both stable and unstable manifolds (the results will be given elsewhere). A possibility of generalization of the method to other systems needs further investigations.

## ACKNOWLEDGEMENTS

The work was supported in part by the Russian Foundation for Basic Investigations (Grants No. 99-02-18315).